\documentclass[superscriptaddress,twocolumn,showpacs,preprintnumbers,amsmath,amssymb,pra,10pt,aps]{revtex4-1}

\usepackage{graphicx}
\usepackage{dcolumn}
\usepackage{bm}
\usepackage{subfigure}
\usepackage{multirow}
\usepackage{soul}
\usepackage{float}
\usepackage[normalem]{ulem}
\usepackage[usenames,dvipsnames]{color}
\usepackage{bbold}
\usepackage{hyperref}

\begin{document}

\title{Local-moment magnetism in Mn-based pnictides}
\author{Matteo Crispino}
\affiliation{Institut für Theoretische Physik und Astrophysik and Würzburg-Dresden Cluster of Excellence ct.qmat, Universität Würzburg, 97074 Würzburg, Germany}
\author{Niklas Witt}
\affiliation{Institut für Theoretische Physik und Astrophysik and Würzburg-Dresden Cluster of Excellence ct.qmat, Universität Würzburg, 97074 Würzburg, Germany}
\author{Tommaso Gorni}
\altaffiliation{Present address: CINECA National Supercomputing Center, Casalecchio di Reno, I-40033 Bologna, Italy}
\affiliation{LPEM, ESPCI Paris, PSL Research University, CNRS, Sorbonne Universit\'e, 75005 Paris France}
\author{Giorgio Sangiovanni}
\affiliation{Institut für Theoretische Physik und Astrophysik and Würzburg-Dresden Cluster of Excellence ct.qmat, Universität Würzburg, 97074 Würzburg, Germany}
\author{Luca de' Medici}
\affiliation{LPEM, ESPCI Paris, PSL Research University, CNRS, Sorbonne Universit\'e, 75005 Paris France}

\date{\today}
\begin{abstract}
We report a comprehensive study of electronic-correlation effects in Manganese-based antiferromagnetic pnictides BaMn$_2$Pn$_2$ (Pn=P,As,Sb,Bi). Our density functional theory plus slave-spin mean-field simulations indicate that all the compounds lie on the strong-coupling side of an itinerant-to-localized moment crossover, corresponding to the critical interaction strength for the Mott transition in the high-temperature paramagnetic phase. We also show that the experimental Néel temperature of each compound scales with the distance from this crossover.
\end{abstract}


\maketitle
The discovery of high-temperature superconductivity in Fe-based superconductors (FeSCs)\cite{Kamihara_Hosono_First_FeSC} has captured the attention of the community working on strongly-correlated materials due to the plethora of fascinating physics they exhibit: superconductivity at high temperature and under pressure\cite{Sefat_FeSCs-Pressure,Sun_FeSCs-Pressure}, magnetism\cite{Cruz-1111_Magnetism,Huang-Magnetism_122-neutrons}, quantum criticality\cite{Kasahara_Matsuda-Quantum_criticality_BaFeAsP}, and realization of heavy fermion (HF) physics in absence of f-electrons\cite{Hardy_BaFe2As2_SC,Hardy_BaFe2As2_SS, Crispino_d-electron-HF}. Their normal, non-superconduting phase has been portrayed as a ”Hund metal”\cite{Georges_Kotliar-Hundmetals_PhysToday}, characterized by a predominant role of Hund’s coupling that favors a strongly correlated state\cite{Haule_Coherence_Incoherence}, mostly with orbitally-differentiated correlations\cite{deMedici_Hassan_OSMT,Yin_Haule_Kotliar_FeSC_Magnetizations,Yu_Si_U(1)_Slave_Spin, deMedici_Giovannetti_Capone}.
One of the main classes of these compounds is the so-called "122 family" of iron pnictides (FePns), where BaFe$_2$As$_2$ stands out as a parent compound that can be chemically modulated in several ways, showing a rich phase diagram in particular upon carrier doping\cite{Avci_BaFe2As2_PD}. The latter turns the (orthorhombically distorted) antiferromagnetic metal found at stoichiometry into a (tetragonal) paramagnetic one, and at low temperatures gives rise to a superconducting dome peaking at $\sim$38K on the hole-doped side.  
Indeed in BaFe$_2$As$_2$, where conduction bands consist mainly of the five Fe 3d-orbitals and host nominally 6 electrons per Fe atom at stoichiometry, hole-doping can be obtained by substituting Ba atoms with alkali-metals (K, Rb or Cs), thus introducing half a hole per Fe atom. 
Along with its influence on the superconducting dome, hole-doping changes substantially also the physics of the normal phase. It enhances correlations and their orbital selectivity\cite{deMedici_Giovannetti_Capone,Hardy_BaFe2As2_SS}, giving rise to the coexistence of heavy and light electrons. This behavior is all the more pronounced the closer the compounds is pushed towards half filling, where 5 electrons would occupy the conduction bands. However, complete alkali substitution can at most reach half the way (i.e. an occupancy of 5.5 electrons/Fe), where already heavy-fermionic behaviour begins to appear\cite{Hardy_BaFe2As2_SS}. Further hole doping can be obtained, by substituting Fe with Cr e.g. in CsFe$_2$As$_2$\cite{Crispino_d-electron-HF}, where it strongly enhances the heavy-fermionic behaviour. Then at high Cr concentrations, the physics abruptly changes, and the system most likely enters an antiferromagnetic (AF) phase (possibly G-type, as it happens for Cr-substituted BaFe$_2$As$_2$\cite{Marty_Competing_Order_BaFe2As2}).

Indeed, on general grounds, a Mott insulating state is expected already at moderate interaction strengths in these 3d-electron systems at half-filling, owing to the strong impact that Hund's coupling has both on the energetic cost of charge fluctuations and in impeding the metallic screening of the large local moment that forms in a half-filled 5-orbital shell\cite{deMedici_Hund_corr,demedici_Janus}. This is confirmed by direct simulations of these materials within density-functional theory (DFT)+dynamical mean-field methods\cite{Ishida_Mott_d5_nFL_Fe-SC,deMedici_Giovannetti_Capone}. However, since the Mott insulator has a finite entropy at zero temperature, an antiferromagnetic symmetry-breaking order is expected to take over at low temperatures\cite{Bascones_gtype_FeSCs,Bascones_Orbital_FeSCs,Misawa_Piedone_Imada,Crispino_SSMF_AFM}.
It is therefore not suprising that BaMn$_2$As$_2$, which has the structure of undistorted BaFe$_2$As$_2$ with Fe replaced by Mn and hosts 5 electrons in the 5 Mn d-orbitals, is experimentally found to be a G-type antiferromagnetic insulator\cite{Singh_BaMn2As2}.
This evidence alone, however, cannot be considered as a smoking gun of the Mott insulating physics at the heart of the scenario for the physics of the FeSCs outlined above, since AF can also be of weak-coupling origin, i.e. a Fermi-surface instability of the Slater type.

In this perspective, it is interesting that several isovalent substitution of the pnictogen in Mn-based compounds are possible, i.e. BaMn$_2$Pn$_2$ with Pn = P, As, Sb, Bi, and that they do not alter the nature of the magnetic order, but only modify the unit cell dimensions\cite{Jacobs_BaMn2Pn2-exp}. This allows for an investigation of the half-filled members of the 122 Mn pnictides (MnPns) family as a function of chemical pressure, possibly allowing to locate the signature of a weak-to-strong coupling crossover, and the position of the compounds on the weak- or strong-coupling side of it.

Experimentally, MnPns have been widely studied thanks to transport and thermoelectric\cite{An_BaMn2As2,Ogasawara_Localization_BaMn2Bi2}, magnetic\cite{Johnston_BaMn2As2_Heisenberg,Huynh_Itinerat-AF_BaMn2Pn2,Ogasawara_Magnetoresistance_BaMn2Bi2}, and optical\cite{McNally_BaMn2As2} measurements. From a theoretical perspective, efforts primarily focused on BaMn$_2$As$_2$ and BaMn$_2$Sb$_2$, which have been investigated using DFT\cite{An_BaMn2As2,Zhang_BaMn2As2-Sb2_ARPES-DFT}, DFT+dynamical mean-field theory (DMFT)\cite{McNally_BaMn2As2,Zingl_BaMn2As2_DFT-DMFT} and Gutzwiller method\cite{Yao_Gutzwiller-BaFeMnAs}, as well as a Heisenberg-model interpretation\cite{Johnston_BaMn2As2_Heisenberg}. There are indications making MnPns fall in the class of Hund-Mott insulators\cite{McNally_BaMn2As2}, with small gap\cite{Jacobs_BaMn2Pn2-exp,Singh_BaMn2As2-TN,Sangeetha_BaMn2Sb2-TN} and prone to form local magnetic moments\cite{An_BaMn2As2,Johnston_BaMn2As2_Heisenberg}.

A comprehensive experimental study of 122 MnPns\cite{Jacobs_BaMn2Pn2-exp} recently observed a monotonic decrease of N\'eel temperature ($T_N$) with increasing tetragonal lattice parameters by pnictogen substitution. Indeed, $T_N=$795K for BaMn$_2$P$_2$\cite{Jacobs_BaMn2Pn2-exp}, 618K for BaMn$_2$As$_2$\cite{Singh_BaMn2As2-TN}, 450K for BaMn$_2$Sb$_2$\cite{Sangeetha_BaMn2Sb2-TN}, and 387.2K for BaMn$_2$Bi$_2$\cite{Saparov_BaMn2Bi2-TN}. 

Here, we present a theoretical study encompassing all the members of the 122 MnPns. By a combination of DFT and many-body methods, we locate the interaction-driven itinerant-to-localized moment crossover (ILMC), and show that the MnPns all lie on the strong-coupling side of it. The proximity to the ILMC strongly influences the $T_N$, as we find that the closer the MnPn is to the ILMC, the higher is its N\'eel temperature. We also provide a mean-field estimate of $T_N$, successfully reproducing the experimentally observed trend\cite{Jacobs_BaMn2Pn2-exp}. 
Our findings therefore support the scenario in which 122 (both Mn and Fe) pnictides are (magnetically ordered) Mott insulators at half filling, which imply the influence of Mott physics on the normal and superconducting phases found in the phase diagram at larger fillings.

This work is organized as follow. In Sec.~\ref{Sec:Model_Method}, we present the model and the method used throughout the paper. In Sec.~\ref{Sec:PM-AF}, we present a characterization of the competing PM and AF phases, which is fundamental to address the ILMC, subject of Sec.~\ref{Sec:ILMC}. Sec.~\ref{Sec:TN} is instead dedicated to the connection between $T_N$ and the ILMC, and to the comparison with experimental results. Sec.~\ref{Sec:Conclusions} summarizes our conclusions and perspectives.

\section{Model and method}\label{Sec:Model_Method}
\begin{table*}[tb]
    \centering
    \caption{Experimentally measured tetragonal lattice parameters and N\'eel temperature, theoretically calculated critical interactions for the Mott transition ($U_c$) and the insurgence of magnetism ($U_m$), and staggered magnetization ($m$) at $U=3\,$eV for BaMn$_2$Pn$_2$ (Pn=P,As,Sb,Bi). All the experimental values are taken from Ref.\cite{Jacobs_BaMn2Pn2-exp} and references within. $U_c$, $U_m$ and $m$ are calculated within DFT+SSMF.}\label{tab:Parameters}
    \begin{tabular}{|c|c|c|c|c|c|c|c|}
    \hline
         & a($\mathrm{\AA}$) & c($\mathrm{\AA}$) & $z_{\mathrm{Pn}}$ & $T_N$(K) & $U_c(eV)$& $U_m(eV)$ & $m(U=3\,eV)$\\
    \hline
     BaMn$_2$P$_2$ & 4.0381 & 13.0653 & 0.3568 & 795 & 2.9 & 1.5 & 4.65 \\
    \hline
     BaMn$_2$As$_2$ & 4.1686 & 13.473 & 0.3615 & 618 & 2.5 & 1.2 & 4.75 \\
    \hline
     BaMn$_2$Sb$_2$ & 4.397 & 14.33 & 0.3642 & 450 & 2.1 & 1.0 & 4.83 \\
    \hline
     BaMn$_2$Bi$_2$ & 4.4902 & 14.687 & 0.3692 & 387.2 & 1.8 & 0.9 & 4.87 \\
    \hline
    \end{tabular}
\end{table*}

We model the four compounds of the BaMn$_2$Pn$_2$ family (Pn = P, As, Sb, Bi) using the experimental crystal structures with space group I4/mmm (taken from Ref.\cite{Jacobs_BaMn2Pn2-exp}, and reported in Tab. ~\ref{tab:Parameters}) through DFT calculations (see details in Appendix~\ref{app:DFT}). For each of them, the conduction bands have been parametrized with a tight-binding model using projection on five maximally-localized Wannier functions centered on each Mn site.

The effects of electronic correlations are addressed through the multi-orbital Hubbard model\cite{Hubbard_Hubbard_Ham,Kanamori_Kanamori_Ham}:
\begin{equation}\label{eqn:Hubbard_Ham}
    \hat{H}=\sum_{i\neq j,mm',\sigma}{t^{mm'}_{ij}\hat{d}^\dagger_{im\sigma}\hat{d}_{jm'\sigma}}+\sum_{im\sigma}{\epsilon_{im\sigma}\hat{n}_{im\sigma}}+\hat{H}_{\mathrm{int}},
\end{equation}
where $\hat{d}^\dagger_{im\sigma}$ $(\hat{d}_{im\sigma})$ creates (destroys) an electron at site $i$ and orbital $m$, with spin $\sigma$; $\hat{n}_{im\sigma}$ is the number operator. As per the tight-binding parametrization $t_{ij}^{mm'}$ are the hopping amplitudes between different sites ($i$ $j$) and orbitals ($m$ $m'$) and $\epsilon_{im}=t_{ii}^{mm}$ are the onsite energies. 
As customarily done\cite{Georges_deMedici_Mravlje_Rev,Hardy_BaFe2As2_SS,Crispino_d-electron-HF}, we assume the interaction Hamiltonian  $\hat{H}_{int}$ in Eq.~\ref{eqn:Hubbard_Ham} to be in the purely-local density-density form:
\begin{align}\label{eqn:density-density-Hint}
    \hat{H}_{int}& = U\sum_{i,m}{\hat{n}_{im\uparrow}\hat{n}_{im\downarrow}}+(U-2J)\sum_{i,m\neq m'}{\hat{n}_{im\uparrow}\hat{n}_{im'\downarrow}} \nonumber \\
    & +(U-3J)\sum_{i,m<m',\sigma}{\hat{n}_{im\sigma}\hat{n}_{im'\sigma}}.    
\end{align}
In Eq.~\ref{eqn:density-density-Hint}, $U$ is the Coulomb repulsion between electrons with opposite spin in the same orbital, while $J$ is the Hund's coupling. To solve the Hubbard Hamiltonian, we use the slave-spin mean field\cite{deMedici_SSMF_OSMT,deMedici_Hassan_OSMT,Yu_Si_U(1)_Slave_Spin,Yu_OSMT-Fe-pnictides}(SSMF) as implemented in Ref.\cite{Crispino_SSMF_AFM}, that presents a generalization of the method allowing to study the broken-symmetry phases we are interested in. More details of the slave-spin formalism and its derivation can be found in Appendix~\ref{app:SSMF} and in Ref.\cite{Crispino_SSMF_AFM}.

We recall that SSMF decouples Eq.~\ref{eqn:Hubbard_Ham} in a system of renormalized non-interacting fermions coupled with a system of interacting (auxiliary) $\frac{1}{2}$-spin (both carrying the same indices as the original fermions). The former is in the following characterized by the operators $\hat{f}_{im\sigma}$, while the latter by the operator $\hat{O}_{im\sigma}=\hat{S}^-_{im\sigma}+c_{im\sigma} \hat{S}^+_{im\sigma}$, where $c_{im\sigma}$ is an arbitrary gauge of the method, chosen in order to retrieve the correct non-interacting limit\cite{deMedici_Hassan_OSMT}. 
The introduction of the auxiliary degrees of freedom enlarges the Hilbert space and generates unphysical states. These can be avoided by enforcing a constraint $\hat{f}^\dagger_{im\sigma}\hat{f}_{im\sigma}=\hat{S}^z_{im\sigma}+\frac{1}{2}$ (in practice this is done on average, through a set of time- and site-independent Lagrange multipliers), leading to the practical advantage that the interaction can now be expressed in terms of $\hat{S}^z_{im\sigma}$ only. 
The resulting total energy of the system can then be easily determined as the sum of three contributions: kinetic energy
\begin{equation}\label{eqn:E_kin}
    E_\mathrm{kin}  = \sum_{ij,mm',\sigma}t^{mm'}_{ij}\sqrt{Z_{im\sigma}Z_{jm'\sigma}}\langle \hat{f}^\dagger_{im\sigma}\hat{f}_{jm'\sigma} \rangle,
\end{equation}
potential energy
\begin{align}\label{eqn:E_pot}
    E_{\mathrm{pot}}& =U\sum_{i,m}{\langle \hat{S}^z_{im\uparrow}\hat{S}^z_{im\downarrow} \rangle}+(U-2J)\sum_{i,m\neq m'}{\langle \hat{S}^z_{im\uparrow}\hat{S}^z_{im'\downarrow} \rangle} \nonumber \\
    & +(U-3J)\sum_{i,m<m',\sigma}{\langle \hat{S}^z_{im\sigma}\hat{S}^z_{im'\sigma} \rangle},
\end{align}
and local energy
\begin{equation}\label{eqn:E_loc}
    E_{\mathrm{loc}} = \sum_{im\sigma} \epsilon_{im\sigma} \langle \hat{n}^f_{im\sigma} \rangle, 
\end{equation}
where $Z_{im\sigma}=|\langle \hat{O}_{im\sigma} \rangle|^2$ are the renormalization factors of the fermionic system due to interaction. We comment here that, contrary to what has been done in other related studies\cite{McNally_BaMn2As2,Zingl_BaMn2As2_DFT-DMFT}, we are considering only the $d$ orbitals of Mn in the construction of the Wannier model, entering through $t^{mm'}_{ij}$ and $\epsilon_{im\sigma}$ in Eq.~\ref{eqn:Hubbard_Ham}. In the context of SSMF, this procedure has been proven to be very effective in characterizing both the effects of electronic correlations and G-type ordering\cite{Crispino_d-electron-HF}.
Moreover, following the results of Refs.~\cite{Hardy_BaFe2As2_SS,Crispino_d-electron-HF}, we expect the values of $U\simeq$3eV and $J/U\simeq$0.15 estimated for the 122 FeSC family by the constrained Random-Phase Approximation (cRPA) to be roughly correct\footnote{$U$=2.8eV and $J$=0.43eV (yielding $J/U\simeq$0.15) are the values calculated in Ref.\cite{Miyake_U_J_used} for BaFe$_2$As$_2$. The Sommerfeld coefficient of the whole 122 family is quite correctly reproduced\cite{Hardy_BaFe2As2_SS} by the present approach with these values. It was found\cite{Crispino_d-electron-HF} that the extremely overdoped compounds are captured even better by slightly increasing    $U\simeq$3eV and the same $J/U$.} also for MnPns. Nevertheless, we are going to study these models as a function of the interaction strength, while keeping the cRPA ratio $J/U$=0.15 in our simulations. All the calculations are performed at zero temperature.
\\

\begin{figure}
    \centering
    \includegraphics[width=0.8\columnwidth]{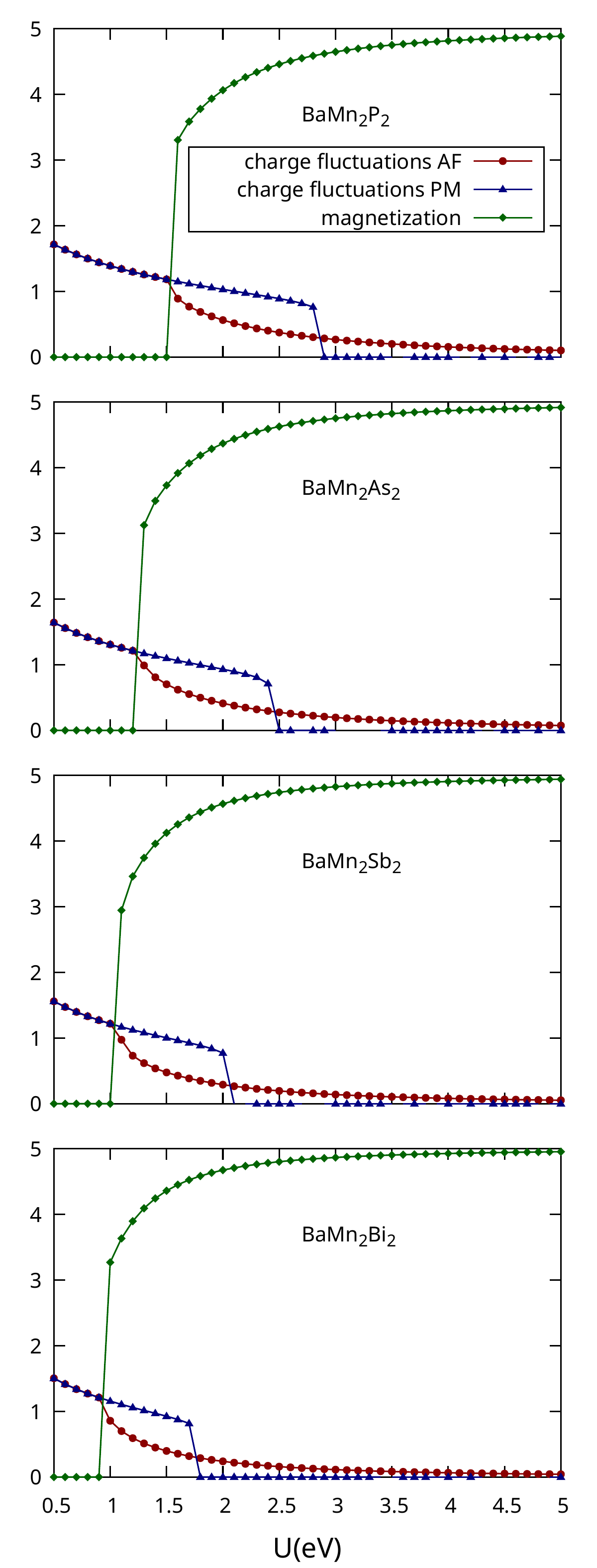}
    \caption{Charge fluctuations for the antiferromagnetic (red circles) and paramagnetic (blue triangles) phase, as a function of the interaction, for BaMn$_2$Pn$_2$ (Pn=P,As,Sb,Bi). Within DFT+SSMF method, the vanishing of the charge fluctuations in the PM corresponds to the Mott transition ($U_c$). The green diamonds report the staggered magnetization (in unit of $\mu_B$) of Mn, becoming finite at the first-order magnetic transition ($U_m$), and rapidly saturating with $U$. Both $U_m$ and $U_c$ decrease along the series of 122 MnPns.}
    \label{fig:Docc-m-All}
\end{figure}

\section{Slave-spin mean field study of PM and AF phases}\label{Sec:PM-AF} 
SSMF is a dynamical mean-field method, and can indeed explore the emergence of symmetry-broken phases while capturing the local many-body correlations in each of the phases encountered. Here, we study the G-type AF phase found experimentally\cite{Jacobs_BaMn2Pn2-exp} along with the paramagnetic phase with unbroken symmetry where it arises. 
Indeed, at small interaction strength $U$ and $J$ the studied systems remain paramagnetic, albeit with increasing hopping renormalization (i.e. the quasiparticle weights - and inverse mass-enhancements - $Z_{im\sigma}$ diminish gradually from unity, which is the value they have for the non-interacting system). We plot the staggered magnetization in Fig.~\ref{fig:Docc-m-All} (green curves), which becomes abruptly nonzero at a (material-dependent) critical interaction strength $U_m$.
Indeed, moving in the series from P to Bi, the compound becomes AF for lower and lower values of interaction $U_m$. The corresponding values of $U_m$ are reported in Table~\ref{tab:Parameters}. We also notice that the staggered magnetization tends to saturate more rapidly going from P to Bi.

Correlations remain weak all along the AF phase, owing to the strong spin polarization that reduces quantum fluctuations. Conversely, they keep increasing in the paramagnetic phase that can be accessed at interactions larger than $U_m$ by forbidding symmetry-broken solution.
In the paramagnetic phase, all the compounds undergo a metal-to-insulator transition (MIT), whose critical value $U_c$ decreases going from P to Bi. The MIT can be clearly seen in Fig.~\ref{fig:Docc-m-All} where we show the charge fluctuations i.e. $\langle \left(\sum_{m\sigma} \hat{n}_{m\sigma} \right)^2 \rangle - \langle \sum_{m\sigma} \hat{n}_{m\sigma}\rangle^2$ as a function of $U$ in the PM phase. Within SSMF, the Mott transition is characterized by the vanishing of the quasiparticle weights $Z_{im\sigma}$, together with the vanishing of the charge fluctuations\cite{Brinkman_Gutzwiller-MIT, kotliar_ruckenstein, Klejnberg-MIT,Fanfarillo_Charge-Fluctuations}. This defines the critical interaction strength for the Mott transition $U_c$ for each compound, the values of which are reported in Table~\ref{tab:Parameters}. 

We observe that $U_c(\textrm{P})$$>$$U_c(\textrm{As})$$>$$U_c(\textrm{Sb})$$>$$U_c(\textrm{Bi})$\footnote{We notice also that the ratio between the critical interaction for the MIT in the PM phase and for the developing of a staggered magnetization is $U_c/U_m\simeq 2$ for all the 122-Mn}, indicating that, at the same $U$, the compound tends to be progressively more localized by pnictogen substitution from P to Bi. Indeed, pnictogen substitution from P to Bi increases the tetragonal lattice parameters, and the further apart the Mn atoms are, due to the progressively wider lattice spacing, the harder it is for electrons to hop, up to a point in which the energy cost due to repulsion is so high to overcome the electronic delocalization. 

For the aforementioned (Sec.~\ref{Sec:Model_Method}) value of the interaction $U=$3eV and $J/U=$0.15, DFT+SSMF simulations in the PM phase place BaMn$_2$P$_2$ slightly above the Mott transition, and all the other compounds progressively further into the Mott-insulating phase.

This ab-initio estimate supports the aforementioned physical scenario, i.e. the doped paramagnetic phase of 122 compounds is influenced by the presence of a Mott insulator at half filling.  The same estimate can be further supported by analyzing the AF phase in comparison with the experiments.

Indeed, the G-type AF symmetry-broken ground state has a different source of stabilization energy compared to the unbroken paramagnetic phase, depending on this being itinerant or localized in nature. In the former case, itinerant antiferromagnetism is accompanied by a gain in potential energy, while in the latter case, localized magnetism is realized owing to a gain in kinetic energy\cite{Taranto-Optics_AF, Crispino_SSMF_AFM}. Indeed, even if the ground state is a G-type AF, the comparison with the competing PM phase at T=0 gives a significant insight into the excited states of the system which differ for an itinerant and a local-moment AF.

This dualism has far-reaching consequences, since the total energy gain, in absolute value, is highest around the interaction strength corresponding to the Mott transition - which then marks the crossover from itinerant to a localized AF - and so is the N\'eel temperature\cite{Mravlje_SrTcO3-AFM}, which can be directly compared with experiments.

\section{Itinerant-to-localized moment crossover}\label{Sec:ILMC}
As mentioned above, in the AF phase Coulomb interaction drives an ILMC\cite{Chatzieleftheriou_It-Loc-Mag,Stepanov_Local-Moment,Zingl_BaMn2As2_DFT-DMFT}, where the itinerant and localized regime can be approximately described respectively by a Slater theory\cite{Slater_Magnetism,Szilva_Magnetism-Review} (at weak coupling) and by an Heisenberg theory\cite{Montorsi-The_Hubbard_Model,Chao_Kinetic,MacDonald_t-U_expansion} (at strong coupling).
In the former, the stabilization energy comes from the potential energy\cite{Taranto-Optics_AF,Gull_Slater-Mott,Schaefer_Fate_Mott,Oles_Hubbard-PD}, which increases with $U$ due to the increase of the magnetic polarization. In the latter, the moment is formed and the stabilization energy comes from the coherent hopping processes made available by the AF orientation of the moments. This kinetic energy gain is of order $\sim t^2/U$ and thus decreases with increasing interaction. Quite naturally, at the crossover between these two regimes one expects a maximum in the energy gain and thus a maximal robustness of the AF phase.

\begin{figure*}[tb]
\begin{center}
\includegraphics[width=\textwidth]{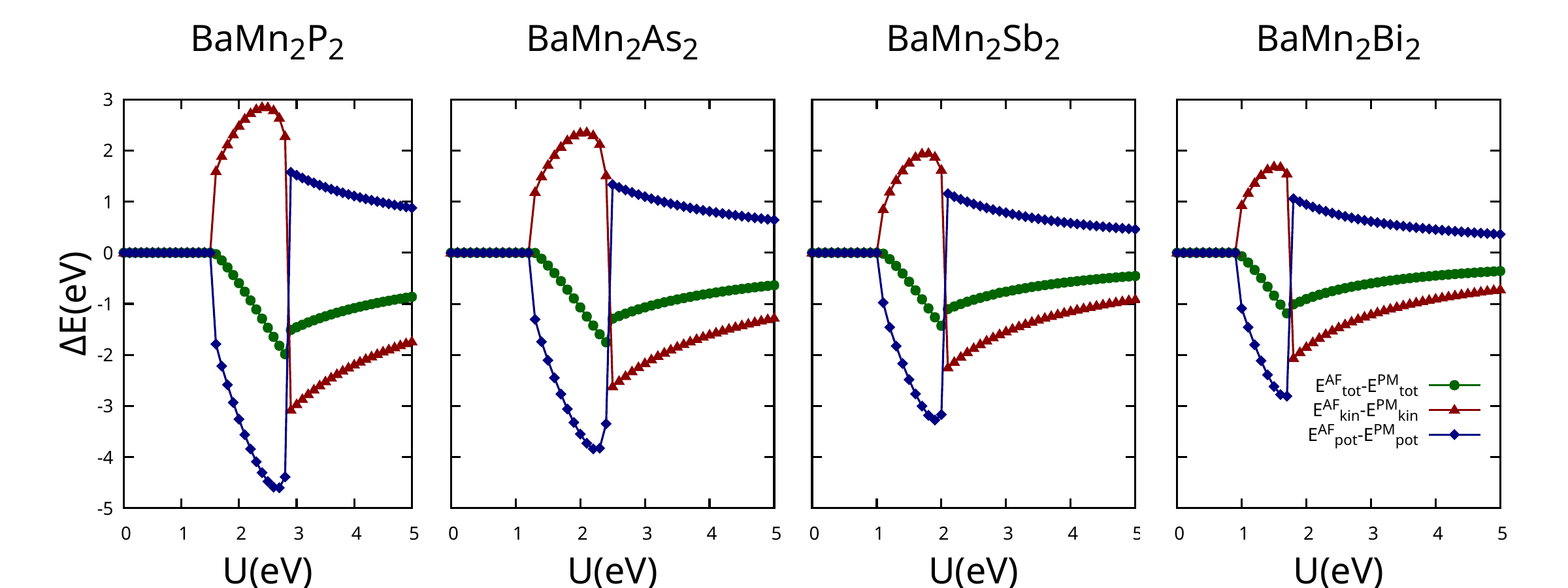}
\caption{Kinetic ($\Delta E_{\mathrm{kin}}= E_{\mathrm{kin}}^{AF}-E_{\mathrm{kin}}^{PM}$), potential ($\Delta E_{\mathrm{pot}}= E_{\mathrm{pot}}^{AF}-E_{\mathrm{pot}}^{PM}$) and total ($\Delta E_{\mathrm{tot}}= E_{\mathrm{tot}}^{AF}-E_{\mathrm{tot}}^{PM}$) energy differences between the AF and PM phases for the family of MnPns. The onset of $\Delta E_\mathrm{tot} \neq 0$ indicates the beginning of the weak-coupling antiferromagnetic phase. The second jump signals the entrance into the local-moment antiferromagnetic region.}\label{fig:Diff_Energy_United}
\end{center}
\end{figure*}

The ILMC can therefore be characterized by looking at the energy of the AF phase with respect to the PM one\cite{Taranto-Optics_AF,Rohringer_Dynamical-Vertex-Approx,Karolak_Ni-Ti_Perovskites,Fratino_Mott-AF}.
The different energy contributions are easily accessible in SSMF and we report in Fig.~\ref{fig:Diff_Energy_United} the difference in the kinetic ($\Delta E_{\mathrm{kin}}=E^{AF}_{\mathrm{kin}}-E^{PM}_{\mathrm{kin}}$) potential ($\Delta E_{\mathrm{pot}}=E^{AF}_{\mathrm{pot}}-E^{PM}_{\mathrm{pot}}$) and total ($\Delta E_{\mathrm{tot}}=E^{AF}_{\mathrm{tot}}-E^{PM}_{\mathrm{tot}}$) energy of the AF with respect to the PM phase, for all the Mn-based pnictides under scrutiny here. 

The transition from PM to AF is signaled by $\Delta E_{\mathrm{tot}} \neq 0$ (i.e. $E^{AF}_{\mathrm{tot}}<E^{PM}_{\mathrm{tot}}$, correctly indicating that the ground state is antiferromagnetic), which corresponds to the onset of the staggered magnetization reported in Fig.~\ref{fig:Docc-m-All}.
There, both $\Delta E_{\mathrm{kin}}$ and $\Delta E_{\mathrm{pot}}$ become finite. 
Then, at weak coupling $\Delta E_{\mathrm{pot}}=E^{AF}_{\mathrm{pot}}-E^{PM}_{\mathrm{pot}}<0$, and overcompensates the positive $\Delta E_{\mathrm{kin}}=E^{AF}_{\mathrm{kin}}-E^{PM}_{\mathrm{kin}}>0$ to obtain the overall $\Delta E_{\mathrm{tot}} < 0$ and stabilize the AF.
At larger interactions, $\Delta E_{\mathrm{tot}}$ displays a jump, which is the consequence of the first-order Mott transition of the paramagnetic phase happening at $U_c$, as reported in Fig.~\ref{fig:Docc-m-All}.
There, $\Delta E_{\mathrm{kin}}$ and $\Delta E_{\mathrm{pot}}$ abruptly cross and the scenario is reversed. The kinetic energy difference becomes negative and overcompensates the now positive potential energy difference. Thus as expected in the strong-coupling regime it is $\Delta E_{\mathrm{kin}}$ that stabilizes the magnetic phase.

We notice here that these conclusions are valid regardless the Pn. Still, $\Delta E_{\mathrm{kin}}$ and $\Delta E_{\mathrm{pot}}$ both decrease in absolute value when going from P to Bi, which implies that the two competing phases are progressively closer in energy.

\section{Proximity to the ILMC and N\'eel temperature}\label{Sec:TN} 
\begin{figure}[tb]
\begin{center}
\includegraphics[width=\columnwidth]{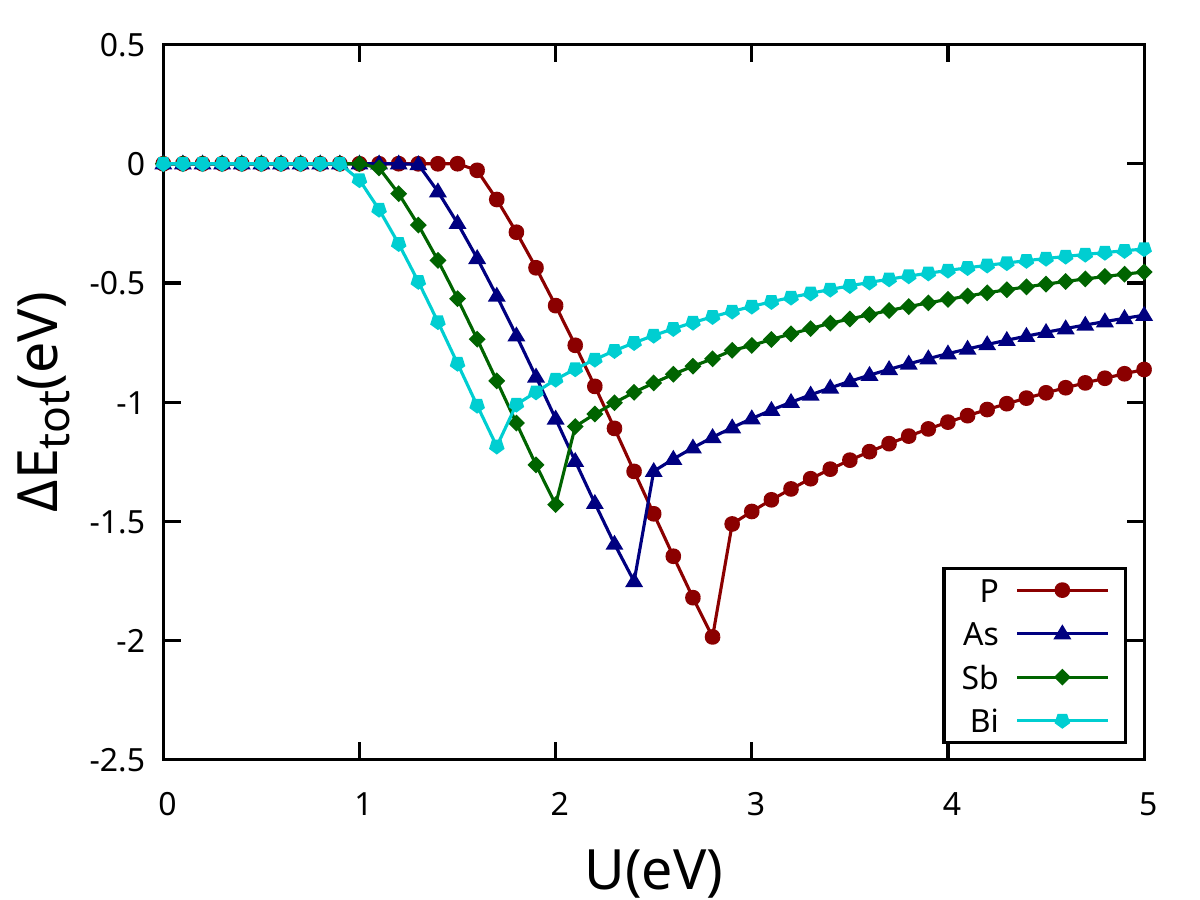}
\caption{Total energy difference between AF and PM phase, for all the studied compounds. Once established, AF is always the ground state. For weak-coupling regime, the compounds display a trend opposite to that of the experimental Néel temperatures\cite{Jacobs_BaMn2Pn2-exp}. The experimental behavior is correctly captured at strong-coupling instead.}\label{fig:all-energy}
\end{center}
\end{figure}

The analysis of the total energy gives a valuable insight in the physics of the compounds also because it allows to establish a link with the experimentally found progression of N\'eel temperature in MnPns. To better visualize this, we collect the total energy differences in Fig.~\ref{fig:all-energy}. For nonzero $\Delta E_{\mathrm{tot}}$, we can easily identify a hierarchy in the robustness of the AF state of the various MnPns, as signaled by the magnitude of $\Delta E_\mathrm{tot}$, which is opposite in the weak and strong coupling, i.e. itinerant and localized, regimes. 

Indeed, for $U > 2.9 \,$eV all the compounds lie beyond the ILMC, and the relation $|\Delta E_{\mathrm{tot}}^{P}| > |\Delta E_{\mathrm{tot}}^{As}| > |\Delta E_{\mathrm{tot}}^{Sb}| > |\Delta E_{\mathrm{tot}}^{Bi}|$ holds systematically for all the studied values of interaction.
This is in fact the same order of the experimental values for $T_N$ reported in Ref.\cite{Jacobs_BaMn2Pn2-exp} (i.e. $T^P_N > T^{As}_N > T^{Sb}_N > T^{Bi}_N$).
The trend is reproduced down to $U=$2.5 eV, thus even if BaMn$_2$P$_2$ is below the ILMC while the other compounds are in the localized-moment regime.

On the other hand for $U<2.5 \,$eV, the hierarchy is different and depend on the precise value of interaction, until $U$ is low enough for all the compounds to be in the itinerant regime, where $|\Delta E_{\mathrm{tot}}^{P}| < |\Delta E_{\mathrm{tot}}^{As}| < |\Delta E_{\mathrm{tot}}^{Sb}| < |\Delta E_{\mathrm{tot}}^{Bi}|$ and the trend is reversed. 

To corroborate these findings, we calculated the trace of the non-interacting static local spin susceptibility tensor $\mathrm{Tr}\chi^0=\sum_{lm}\chi^0_{lmlm}$, where $\chi^0_{lmlm}$ is the bubble of non-interacting Green's functions. $\chi^0$ is usually analyzed for weak-coupling instabilities in RPA(-like) approaches. We used the sparse-sampling approach\cite{Li_Sparse-Sampling,Witt_Fluctuations-SC} of the intermediate representation basis\cite{Shinaoka_Chi-IR,Shinaoka_Chi_Sparse-IR,Wallerberger_Chi-Sparse-IR} to efficiently evaluate the irreducible susceptibility $\chi^0$ on Matsubara frequencies. The trace of the local spin susceptibility is 0.489 eV$^{-1}$, 0.573 eV$^{-1}$, 0.716 eV$^{-1}$ and 0.797 eV$^{-1}$ for P, As, Sb and Bi respectively and the corresponding maximum eigenvalues of $\chi^0$ are 1.185 eV$^{-1}$, 1.379 eV$^{-1}$, 1.820 eV$^{-1}$ and 1.974 eV$^{-1}$. Both the $\chi^0$ analysis and DFT+SSMF simulations then describe the same weak-coupling scenario, opposite to the experimental conclusions.

The hierarchy in the experimental Néel temperatures is thus only captured if the MnPns are all (or all but BaMn$_2$P$_2$ which might be around the ILMC) local-moment antiferromagnets.

In order to make this qualitative claim more quantitative, we can use the simplest estimate of $T_N$ in the strong-coupling regime, which can be easily obtained in a Weiss mean-field of the Heisenberg model\cite{Blundell_Magnetism-Condensed-Matter,Johnston_BaMn2As2_Heisenberg}, which yields $T_N = \Delta E_{\mathrm{tot}}/3k_B$ (where $k_B$ is the Boltzmann constant). The proportionality of $T_N$ on $\Delta E_{\mathrm{tot}}$ explicitly illustrates that both these quantities can be taken as a measure of the robustness of the AF phase.
And albeit the $T_N$ estimated in such a simplified framework is expected - and found - to be a gross overestimate, and that for any reasonable value of the interaction (in all cases it is in the thousands rather than hundreds of Kelvin), it is striking that the relative AF robustness within the series of compounds is perfectly captured instead.

Indeed, if we fix the interaction to the ab-initio estimate $U=3\,$eV, we find the ratios between the $T_N$ to match very well the experimental ones. The results are shown in Table~\ref{tab:T_Comparison}, using $T_N$ of BaMn$_2$Bi$_2$ ($T_N^{(\textrm{Bi})}$) for normalization.

\begin{table}[t]
    \centering
    \caption{Progression of N\'eel temperature across the BaMn$_2$Pn$_2$ family (Pn=P,As,Sb,Bi). All the values are normalized with respect to $T_N^{(\mathrm{Bi})}$ as calculated in this work for theoretical simulations and as reported in Ref.\cite{Saparov_BaMn2Bi2-TN} for the experimental values.}\label{tab:T_Comparison}
    \begin{tabular}{|c|c|c|c|}
    \hline
         & $T_N/T_N^{(\mathrm{Bi})}$ (this work) & $T_N/T_N^{(\mathrm{Bi})}$ (experimental) \\
    \hline
     BaMn$_2$P$_2$ & 2.43 & 2.05 (Ref.\cite{Jacobs_BaMn2Pn2-exp}) \\
    \hline
     BaMn$_2$As$_2$ & 1.77 & 1.60 (Ref.\cite{Singh_BaMn2As2-TN}) \\
    \hline
     BaMn$_2$Sb$_2$ & 1.27 & 1.16 (Ref.\cite{Sangeetha_BaMn2Sb2-TN}) \\
    \hline
     BaMn$_2$Bi$_2$ & 1 & 1 (Ref.\cite{Saparov_BaMn2Bi2-TN}) \\
    \hline
    \end{tabular}
\end{table}

Let us note that, as said, even if the relative robustness of the AF state between the various compounds in the series is correctly captured, the sheer value of $T_N$ is grossly overestimated in our framework. There are several well understood reasons for this. First, as already said using the Weiss mean-field relation between $T_N$ and $\Delta E_{\mathrm{tot}}$ for the Heisenberg model is a rough estimate. In any case, any (even dynamical) mean-field neglects non-local correlations yielding an overestimate of $T_N$\cite{Karolak_Ni-Ti_Perovskites,Rohringer_Spatial-Fluctuations}.
Indeed, DFT+DMFT simulations of BaMn$_2$As$_2$ were shown to require a reduction factor of $\sim 2$ for $T_N$ in order to have agreement with experiments\cite{Zingl_BaMn2As2_DFT-DMFT}. 
Second, our SSMF oversimplifies the description of the Mott phase, with respect to which we calculate $\Delta E_{tot}$ at strong coupling. There, the electrons are perfectly localized and the charge fluctuations vanishes like the quasiparticle weight ($Z_{im\sigma}$). Thus $E_\mathrm{kin}=0$ in this phase (see Eq.~(\ref{eqn:E_kin}) with $Z_{im\sigma}=0$), the residual kinetic energy (corresponding to the incoherent excitations in the Hubbard bands) is neglected.
It is also noteworthy that the simplified density-density form of the interaction Eq.~(\ref{eqn:density-density-Hint}) most likely leads to an overestimate of the magnetic contribution to the energy\cite{Crispino_SSMF_AFM}.

Nevertheless, we can conclude that even in our simplified approach the comparison between the theoretical results and experiments unambiguously places all the MnPns (but perhaps BaMn$_2$P$_2$) in the local-moment AF regime on the strong-coupling side of the ILMC. Moreover one realizes that the smaller the ligand ionic size is (which implies smaller lattice parameters, corresponding to a negative chemical pressure, placing the system closer to the itinerant-to-localized crossover), the higher the N\'eel temperature is\cite{Zingl_BaMn2As2_DFT-DMFT,Mravlje_SrTcO3-AFM}.

\section{Conclusions}\label{Sec:Conclusions}
We presented a density-functional theory+slave-spin mean-field study of the manganese-based pnictides BaMn$_2$Pn$_2$ (Pn=P,As,Sb,Bi). Our analysis of the paramagnetic and G-type ordered antiferromagnetic phases shows that these compounds tend to form local-moment antiferromagnets. Thanks to the computationally efficiency of SSMF in calculating the total energy of the system, we compared it between the two phases.  The results show, as a function of the interaction strength, a crossover from an itinerant regime to a localized one, stabilized respectively by potential and kinetic energy. 
We conclude that all compounds lie on the strong-coupling side of it. As a direct consequence, the smaller the lattice parameters, the closer the compound is to the itinerant-to-localized moment crossover, and consequentially the higher is the N\'eel temperature. 
A strongly-correlated electron picture, and specifically the proximity to the aforementioned crossover, is fundamental to determine the correct behavior of the family, since in the weakly interacting limit the results are opposite to the experimental evidences. Both DFT+SSMF and local spin susceptibility calculations support this conclusion.  This study could be a fruitful starting point to explore similar compounds as well as doped Mn-based pnictides. We also underline how similar analysis of the energy have been applied to competing pairing mechanism in superconductivity\cite{Garg_BCS-BEC,Kyung_BCS-BEC,Toschi_BCS-BEC}.   

\section*{Aknowledgements} 
M. C. thanks E. Stepanov and M. D\"urrnagel for fruitful discussions. M.C. and L. d M. thank L. Fratino, S. Vlaic, and S. Pons for their valuable comments on the energy of the antiferromagnetic phase.
M. C. acknowledge support from ct.qmat (Project‐ID 390858490). N. W. acknowledges support from SFB 1170 (“To-
cotronics”, project No. 258499086). G. S. acknowledges support from the Deutsche Forschungsgemeinschaft (DFG, German Science Foundation) through Project Nr. 468199700.

\appendix

\section{Density-functional theory calculations and tight-binding parametrizations}\label{app:DFT}
To model the four compounds of the BaMn$_2$Pn$_2$ family (Pn = P, As, Sb, Bi) the experimental crystal structures with space group I4/mmm have been taken from Ref.\cite{Jacobs_BaMn2Pn2-exp} (see also Tab. ~\ref{tab:Parameters}). 
DFT calculations with the Perdew-Burke-
Ernzerhof (PBE) functional have been carried out with the Quantum ESPRESSO package\cite{Giannozzi_Quantum-Espresso,Giannozzi_Quantum-Espresso-Advanced}, using norm-conserving pseudopotentials from the PseudoDojo library\cite{vanSetten_PseudoDojo} and a plane-wave cutoff of 90 Ry. Integrals over the Brillouin zone have been converged with a gaussian smearing of $\sigma = 0.01$ Ry and a $\Gamma$-centered uniform grid of 8 × 8 × 8 points.
A tight-binding model is built for each compound by projecting the low-lying PBE Bloch states on a set of five maximally-localized Wannier functions centered on each Mn site, using the Wannier90 code\cite{Pizzi_Wannier90}. The optimal set of the disentanglement and frozen energy windows are reported in Tab.~\ref{tab:TB_model} for each compound. The above procedure results in a tight-binding model for the Mn sites faithfully reproducing the DFT(PBE) low-energy bands.

\begin{table}[t]
    \centering
    \caption{Disentanglement and frozen energy windows in eV, with respect to Fermi energy, used in this work to devise a tight-binding model for each compound.}\label{tab:TB_model}
    \begin{tabular}{|c|c|c|c|c|}
    \hline
         & dis\_win\_min & dis\_froz\_min & dis\_froz\_max & dis\_win\_max \\
    \hline
     BaMn$_2$P$_2$ & -2.0835 & -1.3835 & 0.0165 & 2.9165 \\
    \hline
     BaMn$_2$As$_2$ & -3.0357 & -0.9357 & 1.4643 & 2.3643 \\
    \hline
     BaMn$_2$Sb$_2$ & -2.7672 & -0.7672 & 1.5328 & 2.2328 \\
    \hline
     BaMn$_2$Bi$_2$ & -2.8055 & -0.8055 & 1.1945 & 2.1945 \\
    \hline
    \end{tabular}
\end{table}

\section{Slave-spin mean-field equations}\label{app:SSMF}
In the SSMF formalism, we associate to each fermionic degree of freedom, created by the operator $\hat{d}^\dagger_{im\sigma}$ in Eq.~\ref{eqn:Hubbard_Ham}, both a fermionic degree of freedom ($\hat{f}^\dagger_{im\sigma}$) and the $S^z_{im\sigma}$ component of a (slave) spin-$\frac{1}{2}$ variable. 
This procedure enlarges the Hilbert space, and each state of the original Hilbert space is now expressed as the product of a fermionic state and a slave-spin one. In order to avoid unphysical states\cite{Crispino_SSMF_AFM}, we enforce the constraint
\begin{equation}\label{eqn:constraint}
\hat{f}^\dagger_{im\sigma}\hat{f}_{im\sigma}=\hat{S}^z_{im\sigma}+\frac{1}{2},\quad \forall\,im\sigma
\end{equation}
through a set of site, orbital and spin dependent Lagrange multipliers $\{\lambda_{im\sigma}\}$.
The constraint in Eq.~\ref{eqn:constraint} allows to count the number of physical fermions either through $\hat{n}^{f}_{im\sigma}=\hat{f}^\dagger_{im\sigma}\hat{f}_{im\sigma}$  or the associated slave spin, being the state occupied (empty) if $S^z_{im\sigma}=\frac{1}{2}$ ($-\frac{1}{2}$). As a consequence of Eq.~\ref{eqn:constraint}, the interaction in Eq.~\ref{eqn:density-density-Hint} can be expressed as dependent on $\hat{S}^z$ only, i.e. $\hat{H}[\hat{n}^d]\to\hat{H}_{\mathrm{int}}[\hat{S}^z]$. Moreover, in enlarging the Hilbert space, the operator's mapping follows:
\begin{equation}\label{eqn:mapping-operator}
    \hat{d}^{\left( \dagger \right)}_{im\sigma}\to\hat{f}^{\left( \dagger \right)}_{im\sigma}\hat{O}^{\left( \dagger \right)}_{im\sigma} 
\end{equation}
being $\hat{O}_{im\sigma}=\hat{S}^-_{im\sigma}+c_{im\sigma}\hat{S}^+_{im\sigma}$, where $c_{im\sigma}$ is a complex gauge. We choose it to be:
\begin{equation}
    c_{im\sigma}=\frac{1}{\sqrt{\langle \hat{n}^f_{im\sigma} \rangle \left( 1- \langle \hat{n}^f_{im\sigma} \rangle \right) }}-1
\end{equation}
in order to be consistent with the correct non-interacting limit\cite{deMedici_Hassan_OSMT}.
The Hubbard Hamiltonian in Eq.~\ref{eqn:Hubbard_Ham} is then written as:
\begin{align}\label{eqn:Slave-Hubbard-Ham}
    \hat{H}& = \sum_{i\neq j,mm',\sigma}{t^{mm'}_{ij}\hat{O}^\dagger_{im\sigma}\hat{O}_{jm'\sigma}\hat{f}^\dagger_{im\sigma}\hat{f}_{jm'\sigma}}\nonumber \\ 
    & +\sum_{im\sigma}{\epsilon_{im\sigma}\hat{n}^f_{im\sigma}}+\hat{H}_{\mathrm{int}}\left[ \hat{S}^z\right].
\end{align}
The mean-field equations can be derived via a variational approach\cite{Crispino_SSMF_AFM}, under two assumptions. First, we assume the total wave function of the Hamiltonian in Eq.~\ref{eqn:Hubbard_Ham} to be factorizable as the product of a fermionic and slave-spin component. This corresponds to decoupling the fermionic system from the slave-spin one. As a consequence, the constraint in Eq.~\ref{eqn:constraint} can be treated only on average. Secondly, we assume the total slave-spin wave function to be factorizable as a product of single-site slave-spin wave functions. Physically, this is equivalent to a a mean-field for the slave-spin degrees of freedom, in which only one site is considered and all the other act as a mean field on it. Under these assumptions, the original Hamiltonian in Eq.~\ref{eqn:Hubbard_Ham} results in two mean-field, self-consistently coupled, Hamiltonian problems:
\begin{align}\label{eqn:H_S}
    \hat{H}_S& = \sum_{im\sigma}\left( h_{im\sigma} \hat{O}_{im\sigma} + \mathrm{h.c.} \right) \nonumber \\
    & + \sum_{im\sigma} \lambda_{im\sigma} \left( \hat{S}^z_{im\sigma} +\frac{1}{2} \right)+ \hat{H}_{\mathrm{int}} \left [ \hat{S}^z \right]
\end{align}
and 
\begin{align}\label{eqn:H_f}
    \hat{H}_f & = \sum_{i\neq j,mm',\sigma}t^{mm'}_{ij} \sqrt{Z_{im\sigma}Z_{jm'\sigma}}\hat{f}^\dagger_{im\sigma}\hat{f}_{jm'\sigma} \nonumber \\
    & + \sum_{im\sigma}\left( \epsilon_{im\sigma}-\lambda_{im\sigma}+\lambda^0_{im\sigma} \right)\hat{n}^f_{im\sigma}
\end{align}
describing respectively a system of interacting (slave) spins and one of non-interacting fermions, renormalized by the interaction via $Z_{im\sigma}\equiv |\langle \hat{O}_{im\sigma}\rangle|^2$. In Eq.~\ref{eqn:H_S}, $h_{im\sigma}=\sum_{jm'}t^{mm'}_{ij}\langle \hat{O}_{jm'\sigma} \rangle \langle \hat{f}^\dagger_{im\sigma} \hat{f}_{jm'\sigma} \rangle$ acts as a transverse field for the spin system. The term $\lambda^0_{im\sigma}$ in Eq.~\ref{eqn:H_f} comes, in the variational description, from the dependence of $c_{im\sigma}$ on $\langle \hat{n}^f_{im\sigma}\rangle$. It reads as:
\begin{equation}\label{eqn:lambda_0}
    \lambda^0_{im\sigma}=4h_{im\sigma}\sqrt{{Z}_{im\sigma}} \frac{2\langle \hat{n}^f_{im\sigma} \rangle - 1}{4\langle \hat{n}^f_{im\sigma} \rangle \left( 1 - \langle \hat{n}^f_{im\sigma} \rangle \right)}.
\end{equation}
The energy in Sec.~\ref{Sec:Model_Method} is obtained by averaging Eq.~\ref{eqn:Slave-Hubbard-Ham} under the aforementioned variational ansatz and with $Z_{im\sigma}\equiv |\langle \hat{O}_{im\sigma}\rangle|^2$. We underline that, even if $\lambda_{im\sigma}$ and $\lambda^0_{im\sigma}$ do not contribute explicitly to the energy, they nonetheless affect the wave function with respect to which the energy is calculated. 

\bibliography{Bib/bibmc,Bib/bibldm,Bib/FeSC,Bib/publdm}

\end{document}